\documentclass{article}
\usepackage{authblk}

\newcommand{\footurl}[1]{\footnote{\texttt{#1}}}
\newcommand{\term}[1]{\emph{#1}}

\newcommand{\practicesection}[2]{\section{#1}\label{#2}}
\newcounter{PracticeIdentifier}[section]
\newcommand{\practice}[1]{\stepcounter{PracticeIdentifier}\textbf{\emph{{#1}~(\arabic{section}.\arabic{PracticeIdentifier})}}}

\title{Best Practices for Scientific Computing}
\date{\today}

\author[a]{Greg~Wilson}
\author[b]{D.A.~Aruliah}
\author[c]{C.~Titus~Brown}
\author[d]{Neil~P.~Chue~Hong}
\author[e]{Matt~Davis}
\author[f]{Richard~T.~Guy}
\author[g]{Steven~H.D.~Haddock}
\author[h]{Kathryn~D.~Huff}
\author[i]{Ian~M.~Mitchell}
\author[j]{Mark~D.~Plumbley}
\author[k]{Ben~Waugh}
\author[l]{Ethan~P.~White}
\author[m]{Paul~Wilson}

\affil[a]{\small Software Carpentry / gvwilson@software-carpentry.org}
\affil[b]{\small University of Ontario Institute of Technology / Dhavide.Aruliah@uoit.ca}
\affil[c]{\small Michigan State University / ctb@msu.edu}
\affil[d]{\small Software Sustainability Institute / N.ChueHong@epcc.ed.ac.uk}
\affil[e]{\small Space Telescope Science Institute / mrdavis@stsci.edu}
\affil[f]{\small University of  Toronto / guy@cs.utoronto.ca}
\affil[g]{\small Monterey Bay Aquarium Research Institute / steve@practicalcomputing.org}
\affil[h]{\small University of  California / huff@berkeley.edu}
\affil[i]{\small University of British Columbia / mitchell@cs.ubc.ca}
\affil[j]{\small Queen Mary University of London  / mark.plumbley@eecs.qmul.ac.uk}
\affil[k]{\small University College London / b.waugh@ucl.ac.uk}
\affil[l]{\small Utah State University  / ethan@weecology.org}
\affil[m]{\small University of  Wisconsin / wilsonp@engr.wisc.edu}

\begin{document}

\maketitle

\begin{abstract}
Scientists spend an increasing amount of time building and using
software. However, most scientists are never taught how to do this
efficiently. As a result, many are unaware of tools and practices that
would allow them to write more reliable and maintainable code with
less effort. We describe a set of best practices for scientific
software development that have solid foundations in research and
experience, and that improve scientists' productivity and the
reliability of their software.
\end{abstract}

Software is as important to modern scientific research as telescopes and test
tubes. From groups that work exclusively on computational problems, to
traditional laboratory and field scientists, more and more of the daily
operation of science revolves around developing
new algorithms, managing and analyzing the large amounts of data that are
generated in single research projects, and combining disparate
datasets to assess synthetic problems, and other computational tasks.

Scientists typically develop their own software for these purposes
because doing so requires substantial domain-specific knowledge. As a
result, recent studies have found that scientists typically spend 30\%
or more of their time developing software
\cite{hannay2008,prabhu2011}.  However, 90\% or more of them are
primarily self-taught \cite{hannay2008,prabhu2011}, and therefore lack
exposure to basic software development practices such as writing
maintainable code, using version control and issue trackers, code
reviews, unit testing, and task automation.

We believe that software is just another kind of experimental
apparatus \cite{vardi2010} and should be built, checked, and used as
carefully as any physical apparatus.  However, while most scientists
are careful to validate their laboratory and field equipment, most do
not know how reliable their software is
\cite{hatton1994,hatton1997}. This can lead to serious errors
impacting the central conclusions of published research
\cite{merali2010}: recent high-profile retractions, technical
comments, and corrections because of errors in computational methods
include papers in \emph{Science} \cite{chang2006,ferrari2013},
\emph{PNAS} \cite{ma2007}, the \emph{Journal of Molecular Biology}
\cite{chang2007}, \emph{Ecology Letters} \cite{lees2007,currie2007},
the \emph{Journal of Mammalogy} \cite{kelt2008}, \emph{Journal of the
  American College of Cardiology} \cite{jaccretract2013},
\emph{Hypertension} \cite{hypertension2012}
and \emph{The American Economic Review} \cite{herndon2013}.

In addition, because software is often used for more than a single
project, and is often reused by other scientists, computing errors can
have disproportionate impacts on the scientific process. This type of
cascading impact caused several prominent retractions when an error
from another group's code was not discovered until after publication
\cite{merali2010}.  As with bench experiments, not everything must be
done to the most exacting standards; however, scientists need to be
aware of best practices both to improve their own approaches and for
reviewing computational work by others.

This paper describes a set of practices that are easy to adopt and
have proven effective in many research settings.  Our recommendations
are based on several decades of collective experience both building
scientific software and teaching computing to scientists
\cite{aranda2012,wilson2006b}, reports from many other groups
\cite{heroux2009,kane2003,kane2006,killcoyne2009,matthews2008,pitt-francis2008,pouillon2010},
guidelines for commercial and open source software development
\cite{spolsky2000,fogel2005}, and on empirical studies of scientific
computing \cite{carver2007,kelly2009,segal2005,segal2008a} and
software development in general (summarized in \cite{oram2010}). None
of these practices will guarantee efficient, error-free software
development, but used in concert they will reduce the number of errors
in scientific software, make it easier to reuse, and save the authors
of the software time and effort that can used for focusing on the
underlying scientific questions.

For reasons of space,
we do not discuss the equally important (but independent) issues of
reproducible research,
publication and citation of code and data,
and open science.
We do believe,
however,
that all of these will be much easier to implement
if scientists have the skills we describe.

\subsection*{Acknowledgments}

We are grateful to
Joel Adamson,
Aron Ahmadia,
Roscoe Bartlett,
Erik Bray,
Steven Crouch,
Michael Jackson,
Justin Kitzes,
Adam Obeng,
Karthik Ram,
Yoav Ram,
and Tracy Teal
for feedback on this paper.

Neil Chue Hong was supported by the UK Engineering and Physical
Sciences Research Council (EPSRC) Grant EP/H043160/1 for the UK
Software Sustainability Institute.

Ian M. Mitchell was supported by NSERC Discovery Grant \#298211.

Mark Plumbley was supported by EPSRC through a Leadership Fellowship
(EP/G007144/1) and a grant (EP/H043101/1) for SoundSoftware.ac.uk.

Ethan White was supported by a CAREER grant from the US National
Science Foundation (DEB 0953694).

Greg Wilson was supported by a grant from the Sloan Foundation.

\practicesection{Write programs for people, not computers.}{cognition}

Scientists writing software need to write code that both executes
correctly and can be easily read and understood by other
programmers (especially the author's future self).
If software cannot be easily read and understood, it is
much more difficult to know that it is actually doing what it is intended
to do.
To be productive,
software developers must therefore take several aspects of human cognition into account:
in particular,
that human working memory is limited, human pattern
matching abilities are finely tuned, and human attention span is short
\cite{baddeley2009,hock2008, letovsky1986,binkley2009,robinson2005}.

First, \practice{a program should not require its readers to hold
  more than a handful of facts in memory at once}.  Human working
memory can hold only a handful of items at a time, where each item is
either a single fact or a ``chunk'' aggregating several facts
\cite{baddeley2009,hock2008}, so programs should limit
the total number of items to be remembered to accomplish a task.
The primary way to accomplish this is to break programs up
into easily understood functions, each of which conducts a single,
easily understood, task. This serves to make each piece of the
program easier to understand in the same way that breaking up a scientific
paper using sections and paragraphs makes it easier to read. For
example, a function to calculate the area of a rectangle can be
written to take four separate coordinates:

\begin{small}
\begin{verbatim}
def rect_area(x1, y1, x2, y2):
    ...calculation...
\end{verbatim}
\end{small}

\noindent
or to take two points:

\begin{small}
\begin{verbatim}
def rect_area(point1, point2):
    ...calculation...
\end{verbatim}
\end{small}

\noindent
The latter function is significantly easier for people to read and
remember, while the former is likely to lead to errors, not least
because it is possible to call the original with values in the wrong
order:

\begin{small}
\begin{verbatim}
surface = rect_area(x1, x2, y1, y2)
\end{verbatim}
\end{small}

Second, scientists should \practice{make names consistent,
  distinctive, and meaningful}. For example, using non-descriptive
names, like \texttt{a} and \texttt{foo}, or names that are very
similar, like \texttt{results} and \texttt{results2}, is likely to
cause confusion.

Third, scientists should \practice{make code style and formatting
  consistent}. If different parts of a scientific paper used different
formatting and capitalization, it would make that paper more difficult
to read. Likewise, if different parts of a program are indented
differently, or if programmers mix \texttt{CamelCaseNaming} and
\texttt{pothole\_case\_naming}, code takes longer to read and readers
make more mistakes \cite{letovsky1986,binkley2009}.

\practicesection{Let the computer do the work.}{automation}

Science often involves repetition of computational tasks such as
processing large numbers of data files in the same way or regenerating
figures each time new data is added to an existing analysis. Computers
were invented to do these kinds of repetitive tasks but, even today,
many scientists type the same commands in over and over again or click
the same buttons repeatedly \cite{aranda2012}. In addition to wasting
time, sooner or later even the most careful researcher will lose focus
while doing this and make mistakes.

Scientists should therefore \practice{make the computer repeat tasks}
and \practice{save recent commands in a file for re-use}. For example,
most command-line tools have a ``history'' option that lets users
display and re-execute recent commands, with minor edits to filenames
or parameters.  This is often cited as one reason command-line
interfaces remain popular \cite{ray2009,haddock2010}: ``do this
again'' saves time and reduces errors.

A file containing commands for an interactive system is often called a
\term{script}, though there is real no difference between this and a
program.  When these scripts are repeatedly used in the same way, or
in combination, a workflow management tool can be used.  The
paradigmatic example is compiling and linking programs in languages
such as Fortran, C++, Java, and C\# \cite{dubois2003b}. The most
widely used tool for this task is probably
Make\footurl{http://www.gnu.org/software/make}, although many
alternatives are now available \cite{smith2011}. All of these allow
people to express dependencies between files, i.e., to say that if A
or B has changed, then C needs to be updated using a specific set of
commands. These tools have been successfully adopted for scientific
workflows as well \cite{fomel2007}.

To avoid errors and inefficiencies from repeating commands manually,
we recommend that scientists \practice{use a build tool to automate
  workflows}, e.g., specify the ways in which intermediate data files
and final results depend on each other, and on the programs that
create them, so that a single command will regenerate anything that
needs to be regenerated.

In order to maximize reproducibility, everything needed to re-create
the output should be recorded automatically in a format that other
programs can read. (Borrowing a term from archaeology and forensics,
this is often called the \term{provenance} of data.)  There have been
some initiatives to automate the collection of this information, and
standardize its format \cite{openprovenance}, but it is already
possible to record the following without additional tools:

\begin{itemize}

\item unique identifiers and version numbers for raw data records
  (which scientists may need to create themselves);

\item unique identifiers and version numbers for programs and
  libraries;

\item the values of parameters used to generate any given output; and

\item the names and version numbers of programs (however small) used
  to generate those outputs.

\end{itemize}

\practicesection{Make incremental changes.}{incremental}

Unlike traditional commercial software developers, but very much like
developers in open source projects or startups, scientific programmers
usually don't get their requirements from customers, and their
requirements are rarely frozen \cite{segal2008a,segal2008b}. In fact,
scientists often \emph{can't} know what their programs should do next
until the current version has produced some results. This challenges
design approaches that rely on specifying requirements in advance.

Programmers are most productive when they \practice{work in small
  steps with frequent feedback and course correction} rather than
trying to plan months or years of work in advance. While the details
vary from team to team, these developers typically work in steps that
are sized to be about an hour long, and these steps are often grouped
in iterations that last roughly one week. This accommodates the
cognitive constraints discussed in Section~\ref{cognition}, and
acknowledges the reality that real-world requirements are constantly
changing. The goal is to produce working (if incomplete) code after
each iteration. While these practices have been around for decades,
they gained prominence starting in the late 1990s under the banner of
\term{agile development} \cite{martin2002,kniberg2007}.

Two of the biggest challenges scientists and other programmers face
when working with code and data are keeping track of changes (and
being able to revert them if things go wrong), and collaborating on a
program or dataset \cite{matthews2008}.  Typical ``solutions'' are to
email software to colleagues or to copy successive versions of it to a
shared folder, e.g., Dropbox\footurl{http://www.dropbox.com}. However,
both approaches are fragile and can lead to confusion and lost work
when important changes are overwritten or out-of-date files are
used. It's also difficult to find out which changes are in which
versions or to say exactly how particular results were computed at a
later date.

The standard solution in both industry and open source is to
\practice{use a version control system} (VCS)
\cite{mcconnell2004,fogel2005}. A VCS stores snapshots of a project's
files in a \term{repository} (or a set of repositories).  Programmers
can modify their working copy of the project at will, then
\term{commit} changes to the repository when they are satisfied with
the results to share them with colleagues.

Crucially, if several people have edited files simultaneously, the VCS
highlights the differences and requires them to resolve any conflicts
before accepting the changes.  The VCS also stores the entire history
of those files, allowing arbitrary versions to be retrieved and
compared, together with metadata such as comments on what was changed
and the author of the changes. All of this information can be
extracted to provide provenance for both code and data.

Many good VCSes are open source and freely available, including
Subversion\footurl{http://subversion.apache.org},
Git\footurl{http://git-scm.com}, and
Mercurial\footurl{http://mercurial.selenic.com}. Many free hosting
services are available as well
(SourceForge\footurl{http://sourceforge.net}, Google
Code\footurl{http://code.google.com},
GitHub\footurl{https://github.com}, and
BitBucket\footurl{https://bitbucket.org} being the most popular).  As
with coding style, the best one to use is almost always whatever your
colleagues are already using \cite{fogel2005}.

Reproducibility is maximized when scientists \practice{put everything
  that has been created manually in version control}, including
programs, original field observations, and the source files for
papers. Automated output and intermediate files can be regenerated at
need. Binary files (e.g., images and audio clips) may be stored in
version control, but it is often more sensible to use an archiving
system for them, and store the metadata describing their contents in
version control instead \cite{noble2009}.

\practicesection{Don't repeat yourself (or others).}{dry}

Anything that is repeated in two or more places is more difficult to
maintain. Every time a change or correction is made, multiple
locations must be updated, which increases the chance of errors and
inconsistencies.  To avoid this, programmers follow the DRY Principle
\cite{hunt1999}, for ``don't repeat yourself'', which applies to both
data and code.

For data, this maxim holds that \practice{every piece of data must
  have a single authoritative representation in the system}. Physical
constants ought to be defined exactly once to ensure that the entire
program is using the same value; raw data files should have a single
canonical version, every geographic location from which data has been
collected should be given an ID that can be used to look up its
latitude and longitude, and so on.

The DRY Principle applies to code at two scales. At small scales,
\practice{modularize code rather than copying and pasting}.  Avoiding
``code clones'' has been shown to reduce error rates
\cite{juergens2009}: when a change is made or a bug is fixed, that
change or fix takes effect everywhere, and people's mental model of
the program (i.e., their belief that ``this one's been fixed'')
remains accurate. As a side effect, modularizing code allows people to
remember its functionality as a single mental chunk, which in turn
makes code easier to understand.  Modularized code can also be more
easily repurposed for other projects.

At larger scales, it is vital that scientific programmers
\practice{re-use code instead of rewriting it}. Tens of millions of
lines of high-quality open source software are freely available on the
web, and at least as much is available commercially. It is typically
better to find an established library or package that solves a problem
than to attempt to write one's own routines for well established
problems (e.g., numerical integration, matrix inversions, etc.).

\practicesection{Plan for mistakes.}{defensive}

Mistakes are inevitable, so verifying and maintaining the validity of
code over time is immensely challenging \cite{grubb2003}. While no
single practice has been shown to catch or prevent all mistakes,
several are very effective when used in combination
\cite{mcconnell2004,dubois2005,sanders2008}.

The first line of defense is \term{defensive programming}.
Experienced programmers \practice{add assertions to programs to check
  their operation} because experience has taught them that everyone
(including their future self) makes mistakes.  An \term{assertion} is
simply a statement that something holds true at a particular point in
a program; as the example below shows, assertions can be used to
ensure that inputs are valid, outputs are consistent, and so
on\footnote{Assertions do not require language support: it is common
  in languages such as Fortran for programmers to create their own
  test-and-fail functions for this purpose.}.

\begin{small}
\begin{verbatim}
def bradford_transfer(grid, point, smoothing):
    assert grid.contains(point),
           'Point is not located in grid'
    assert grid.is_local_maximum(point),
           'Point is not a local maximum in grid'
    assert len(smoothing) > FILTER_LENGTH,
           'Not enough smoothing parameters'
    ...do calculations...
    assert 0.0 < result <= 1.0,
           'Bradford transfer value out of legal range'
    return result
\end{verbatim}
\end{small}

\noindent
Assertions can make up a sizeable fraction of the code in well-written
applications, just as tools for calibrating scientific instruments can make up a
sizeable fraction of the equipment in a lab.  These assertions serve two
purposes. First, they ensure that if something does go wrong, the program will
halt immediately, which simplifies debugging. Second, assertions are
\term{executable documentation}, i.e., they explain the program as well as
checking its behavior. This makes them more useful in many cases than comments
since the reader can be sure that they are accurate and up to date.

The second layer of defense is \term{automated testing}. Automated tests can
check to make sure that a single unit of code is returning correct results
(\term{unit tests}), that pieces of code work correctly when combined
(\term{integration tests}), and that the behavior of a program doesn't change
when the details are modified (\term{regression tests}). These tests are conducted
by the computer, so that they are easy to rerun every time the program is modified.
Creating and managing tests is easier if programmers \practice{use an
  off-the-shelf unit testing library} to initialize inputs, run tests, and
report their results in a uniform way. These libraries are available for all
major programming languages including those commonly used in scientific computing
\cite{xunit,meszaros2007,osherove2009}.

Tests check to see whether the code matches the researcher's expectations of its
behavior, which depends on the researcher's understanding of the problem at hand
\cite{hook2009,kelly2008,oberkampf2010}. For example, in scientific computing,
tests are often conducted by comparing output to simplified cases, experimental
data, or the results of earlier programs that are trusted. Another approach for
generating tests is to \practice{turn bugs into test cases}
by writing tests that trigger a bug that has been found in the code and (once fixed) will prevent
the bug from reappearing unnoticed. In combination these kinds of testing can
improve our confidence that scientific code is operating properly
and that the results it produces are valid. An additional benefit of testing is
that it encourages programmers to design and build code that is testable (i.e.,
self-contained functions and classes that can run more or less independently of
one another). Code that is designed this way is also easier to understand
(Section~\ref{cognition}) and more reusable (Section~\ref{dry}).

Now matter how good ones computational practice is, reasonably complex code will
always initially contain bugs. Fixing bugs that have been identified is often
easier if you \practice{use a symbolic debugger} to track them down. A better
name for this kind of tool would be ``interactive program inspector'' since a
debugger allows users to pause a program at any line (or when some condition is
true), inspect the values of variables, and walk up and down active function
calls to figure out why things are behaving the way they are. Debuggers are
usually more productive than adding and removing print statements or scrolling
through hundreds of lines of log output \cite{zeller2009}, because they allow
the user to see exactly how the code is executing rather than just snapshots of
state of the program at a few moments in time. In other words, the debugger
allows the scientist to witness what is going wrong directly, rather than having
to anticipate the error or infer the problem using indirect evidence.

\practicesection{Optimize software only after it works correctly.}{performance}

Today's computers and software are so complex that even experts find
it hard to predict which parts of any particular program will be
performance bottlenecks \cite{jones1999}. The most productive way to
make code fast is therefore to make it work correctly, determine
whether it's actually worth speeding it up, and---in those cases where
it is---to \practice{use a profiler to identify bottlenecks}.

This strategy also has interesting implications for choice of
programming language. Research has confirmed that most programmers
write roughly the same number of lines of code per unit time
regardless of the language they use \cite{prechelt2010}. Since faster,
lower level, languages require more lines of code to accomplish the
same task, scientists are most productive when they \practice{write
  code in the highest-level language possible}, and shift to low-level
languages like C and Fortran only when they are sure the performance
boost is needed\footnote{Using higher-level languages also helps
  program comprehensibility, since such languages have, in a sense,
  ``pre-chunked'' the facts that programmers need to have in
  short-term memory}. Taking this approach allows more code to be
written (and tested) in the same amount of time.  Even when it is
known before coding begins that a low-level language will ultimately
be necessary, rapid prototyping in a high-level language helps
programmers make and evaluate design decisions quickly. Programmers
can also use a high-level prototype as a test oracle for a
high-performance low-level reimplementation, i.e., compare the output
of the optimized (and usually more complex) program against the output
from its unoptimized (but usually simpler) predecessor in order to
check its correctness.

\practicesection{Document design and purpose, not mechanics.}{embeddoc}

In the same way that a well documented experimental protocol makes
research methods easier to reproduce, good documentation helps people
understand code. This makes the code more reusable and lowers
maintenance costs \cite{mcconnell2004}. As a result, code that is well
documented makes it easier to transition when the graduate students
and postdocs who have been writing code in a lab transition to the
next career phase.  Reference documentation and descriptions of design
decisions are key for improving the understandability of
code. However, inline documentation that recapitulates code is
\emph{not} useful. Therefore we recommend that scientific programmers
\practice{document interfaces and reasons, not implementations}. For
example, a clear description like this at the beginning of a function that
describes what it does and its inputs and outputs is useful:

\begin{small}
\begin{verbatim}
def scan(op, values, seed=None):
    # Apply a binary operator cumulatively to the values given
    # from lowest to highest, returning a list of results.
    # For example, if 'op' is 'add' and 'values' is '[1, 3, 5]',
    # the result is '[1, 4, 9]' (i.e., the running total of the
    # given values).  The result always has the same length as
    # the input.
    # If 'seed' is given, the result is initialized with that
    # value instead of with the first item in 'values', and
    # the final item is omitted from the result.
    # Ex: scan(add, [1, 3, 5], seed=10) => [10, 11, 14]

    ...implementation...
\end{verbatim}
\end{small}

In contrast, the comment in the code fragment below does nothing to
aid comprehension:

\begin{small}
\begin{verbatim}
i = i + 1        # Increment the variable 'i' by one.
\end{verbatim}
\end{small}

If a substantial description of the implementation of a piece of
software is needed, it is better to \practice{refactor code in
  preference to explaining how it works}, i.e., rather than write a
paragraph to explain a complex piece of code, reorganize the code
itself so that it doesn't need such an explanation. This may not
always be possible---some pieces of code simply are intrinsically
difficult---but the onus should always be on the author to convince
his or her peers of that.

The best way to create and maintain reference documentation is to
\practice{embed the documentation for a piece of software in that
software}. Doing this increases the probability that when
programmers change the code, they will update the documentation at the
same time.

Embedded documentation usually takes the form of specially-formatted
and placed comments. Typically, a \term{documentation generator} such
as Javadoc, Doxygen, or
Sphinx\footurl{http://en.wikipedia.org/wiki/Comparison\-\_of\_documentation\_generators}
extracts these comments and generates well-formatted web pages and
other human-friendly documents. Alternatively, code can be embedded in
a larger document that includes information about what the code is
doing (i.e., literate programming). Common approaches to this include
this use of knitr \cite{xie2013knitr} and IPython Notebooks
\cite{perez2007}.

\practicesection{Collaborate.}{collaborate}

In the same way that having manuscripts reviewed by other scientists
can reduce errors and make research easier to understand, reviews of
source code can eliminate bugs and improve readability.  A large body
of research has shown that \term{code reviews} are the most
cost-effective way of finding bugs in code
\cite{fagan1976,cohen2010}. They are also a good way to spread
knowledge and good practices around a team. In projects with shifting
membership, such as most academic labs, code reviews help ensure that
critical knowledge isn't lost when a student or postdoc leaves the
lab.

Code can be reviewed either before or after it has been committed to a
shared version control repository. Experience shows that if reviews
don't have to be done in order to get code into the repository, they
will soon not be done at all \cite{fogel2005}. We therefore recommend
that projects \practice{use pre-merge code reviews}.

An extreme form of code review is \term{pair programming}, in which
two developers sit together while writing code. One (the driver)
actually writes the code; the other (the navigator) provides real-time
feedback and is free to track larger issues of design and consistency.
Several studies have found that pair programming improves productivity
\cite{williams2010}, but many programmers find it intrusive. We
therefore recommend that teams \practice{use pair programming when
  bringing someone new up to speed and when tackling particularly
  tricky problems}.

Once a team grows beyond a certain size, it becomes difficult to keep
track of what needs to be reviewed, or of who's doing what.  Teams can
avoid a lot of duplicated effort and dropped balls if they
\practice{use an issue tracking tool} to maintain a list of tasks to
be performed and bugs to be fixed \cite{dubois2003a}. This helps avoid
duplicated work and makes it easier for tasks to be transferred to
different people. Free repository hosting services like GitHub include
issue tracking tools, and many good standalone tools exist as well,
such as Trac\footurl{http://trac.edgewall.org}.

\section{Conclusion}\label{conclusion}

We have outlined a series of recommended best practices for scientific
computing based on extensive research, as well as our collective
experience.  These practices can be applied to individual work as
readily as group work; separately and together, they improve the
productivity of scientific programming and the reliability of the
resulting code, and therefore the speed with which we produce results
and our confidence in them.  They are also, we believe, prerequisites
for reproducible computational research: if software is not version
controlled, readable, and tested, the chances of its authors (much less anyone else)
being able to re-create results are remote.

Our 25 recommendations are a beginning, not an end.  Individuals and
groups who have incorporated them into their work will find links to
more advanced practices on the Software Carpentry
website\footurl{http://software-carpentry.org/biblio.html}.

Research suggests that the time cost of implementing these kinds of
tools and approaches in scientific computing is almost immediately
offset by the gains in productivity of the programmers involved
\cite{aranda2012}. Even so, the recommendations described above may
seem intimidating to implement.  Fortunately, the different practices
reinforce and support one another, so the effort required is less than
the sum of adding each component separately. Nevertheless, we do not
recommend that research groups attempt to implement all of these
recommendations at once, but instead suggest that these tools be
introduced incrementally over a period of time.

How to implement the recommended practices can be learned from many
excellent tutorials available online or through workshops and classes
organized by groups like Software
Carpentry\footurl{http://software-carpentry.org}. This type of
training has proven effective at driving adoption of these tools in
scientific settings \cite{aranda2012,wilson2013}.

For computing to achieve the level of rigor that is expected
throughout other parts of science, it is necessary for scientists to
begin to adopt the tools and approaches that are known to improve both
the quality of software and the efficiency with which it is
produced. To facilitate this adoption, universities and funding
agencies need to support the training of scientists in the use of
these tools and the investment of time and money in building better
scientific software. Investment in these approaches by both
individuals and institutions will improve our confidence in the
results of computational science and will allow us to make more rapid
progress on important scientific questions than would otherwise be
possible.

\bibliographystyle{plain}
\bibliography{best-practices-scientific-computing-2012}

\begin{thebibliography}{10}

\bibitem{jaccretract2013}
Anon.
\newblock {Retraction notice to “Plasma PCSK9 levels and clinical outcomes in
  the TNT (Treating to New Targets) Trial” [J Am Coll Cardiol
  2012;59:1778–1784}.
\newblock {\em Journal of the American College of Cardiology}, 61(16):1751,
  2013.

\bibitem{aranda2012}
Jorge Aranda.
\newblock {Software Carpentry Assessment Report}.
\newblock http://software-carpentry.org/papers/aranda-assessment-2012-07.pdf,
  2012.

\bibitem{baddeley2009}
Alan Baddeley, Michael~W. Eysenck, and Michael~C. Anderson.
\newblock {\em {Memory}}.
\newblock Psychology Press, 2009.

\bibitem{binkley2009}
D.~Binkley, M.~Davis, D.~Lawrie, and C.~Morrell.
\newblock {To CamelCase or Under\_score}.
\newblock In {\em 2009 IEEE International Conference on Program Comprehension},
  pages 158--167, 2009.

\bibitem{carver2007}
Jeffrey~C. Carver, Richard~P. Kendall, Susan~E. Squires, and Douglass~E. Post.
\newblock {Software Development Environments for Scientific and Engineering
  Software: A Series of Case Studies}.
\newblock In {\em 29th International Conference on Software Engineering}, pages
  550--559, 2007.

\bibitem{chang2007}
Geoffrey Chang.
\newblock {Retraction of 'Structure of MsbA from \emph{Vibrio cholera}: A
  Multidrug Resistance ABC Transporter Homolog in a Closed Conformation' [J.
  Mol. Biol. (2003) 330 419–430]}.
\newblock {\em Journal of Molecular Biology}, 369(2), 2007.

\bibitem{chang2006}
Geoffrey Chang, Christopher~B. Roth, Christopher~L. Reyes, Owen Pornillos,
  Yen-Ju Chen, and Andy~P. Chen.
\newblock {Retraction}.
\newblock {\em Science}, 314(5807):1875, 2006.

\bibitem{cohen2010}
Jason Cohen.
\newblock {Modern Code Review}.
\newblock In Andy Oram and Greg Wilson, editors, {\em Making Software: What
  Really Works, and Why We Believe It}, pages 329--336. O'Reilly, 2010.

\bibitem{currie2007}
David Currie and Jeremy Kerr.
\newblock {Testing, as opposed to supporting, the Mid-domain Hypothesis: a
  response to Lees and Colwell (2007)}.
\newblock {\em Ecology Letters}, 10(9):E9--E10, 2007.

\bibitem{dubois2003a}
P.~Dubois and J.~Johnson.
\newblock {Issue Tracking}.
\newblock {\em Computing in Science \& Engineering}, 5(6):71--77,
  November-December 2003.

\bibitem{dubois2005}
P.~F. Dubois.
\newblock {Maintaining Correctness in Scientific Programs}.
\newblock {\em Computing in Science \& Engineering}, 7(3):80--85, May-June
  2005.

\bibitem{dubois2003b}
P.~F. Dubois, T.~Epperly, and G.~Kumfert.
\newblock {Why Johnny Can't Build (Portable Scientific Software)}.
\newblock {\em Computing in Science \& Engineering}, 5(5):83--88, 2003.

\bibitem{fagan1976}
Michael~E. Fagan.
\newblock {Design and Code Inspections to Reduce Errors in Program
  Development}.
\newblock {\em IBM Systems Journal}, 15(3):182--211, 1976.

\bibitem{ferrari2013}
F.~Ferrari, Y.~L. Jung, P.~V. Kharchenko, A.~Plachetka, A.~A. Alekseyenko,
  M.~I. Kuroda, and P.~J. Park.
\newblock {Comment on "Drosophila Dosage Compensation Involves Enhanced Pol II
  Recruitment to Male X-Linked Promoters"}.
\newblock {\em Science}, 340(6130):273, 2013.

\bibitem{fogel2005}
Karl Fogel.
\newblock {\em {Producing Open Source Software: How to Run a Successful Free
  Software Project}}.
\newblock O'Reilly, 2005.

\bibitem{fomel2007}
S.~Fomel and G.~Hennenfent.
\newblock {Reproducible computational experiments using SCons}.
\newblock In {\em 32nd International Conference on Acoustics, Speech, and
  Signal Processing}, volume~IV, pages 1257--1260, 2007.

\bibitem{grubb2003}
Penny Grubb and Armstrong~A. Takang.
\newblock {\em {Software Maintenance: Concepts and Practice}}.
\newblock World Scientific, 2 edition, 2003.

\bibitem{haddock2010}
Steven Haddock and Casey Dunn.
\newblock {\em {Practical Computing for Biologists}}.
\newblock Sinauer Associates, 2010.

\bibitem{hannay2008}
Jo~Erskine Hannay, Hans~Petter Langtangen, Carolyn MacLeod, Dietmar Pfahl,
  Janice Singer, and Greg Wilson.
\newblock {How Do Scientists Develop and Use Scientific Software?}
\newblock In {\em Second International Workshop on Software Engineering for
  Computational Science and Engineering}, pages 1--8, 2009.

\bibitem{hatton1997}
L.~Hatton.
\newblock {The {T} Experiments: Errors in Scientific Software}.
\newblock {\em Computational Science \& Engineering}, 4(2):27--38, 1997.

\bibitem{hatton1994}
L.~Hatton and A.~Roberts.
\newblock {How Accurate is Scientific Software?}
\newblock {\em IEEE Transactions on Software Engineering}, 20(10):785--797,
  1994.

\bibitem{herndon2013}
T.~Herndon, M.~Ash, and R.~Pollin.
\newblock {Does High Public Debt Consistently Stifle Economic Growth? A
  Critique of Reinhart and Rogoff}.
\newblock Technical report, Political Economy Research Institute, 2013.

\bibitem{heroux2009}
Michael~A. Heroux and James~M. Willenbring.
\newblock {Barely-Sufficient Software Engineering: 10 Practices to Improve Your
  {CSE} Software}.
\newblock In {\em Second International Workshop on Software Engineering for
  Computational Science and Engineering}, pages 15--21, 2009.

\bibitem{hock2008}
Roger~R. Hock.
\newblock {\em {Forty Studies That Changed Psychology: Explorations into the
  History of Psychological Research}}.
\newblock Prentice Hall, 6th edition, 2008.

\bibitem{hook2009}
Daniel Hook and Diane Kelly.
\newblock {Testing for Trustworthiness in Scientific Software}.
\newblock In {\em Second International Workshop on Software Engineering for
  Computational Science and Engineering}, pages 59--64, May 2009.

\bibitem{hunt1999}
Andrew Hunt and David Thomas.
\newblock {\em {The Pragmatic Programmer: From Journeyman to Master}}.
\newblock Addison-Wesley, 1999.

\bibitem{hypertension2012}
Hypertension.
\newblock {Notice of Retraction}.
\newblock {\em Hypertension}, 2012.

\bibitem{jones1999}
Michael~B. Jones and John Regehr.
\newblock {The Problems You're Having May Not Be the Problems You Think You're
  Having: Results from a Latency Study of Windows NT}.
\newblock In {\em 7th Workshop on Hot Topics in Operating Systems}, pages
  96--101, 1999.

\bibitem{juergens2009}
Elmar Juergens, Florian Deissenboeck, Benjamin Hummel, and Stefan Wagner.
\newblock {Do Code Clones Matter?}
\newblock In {\em 31st International Conference on Software Engineering}, pages
  485--495, 2009.

\bibitem{kane2003}
David Kane.
\newblock {Introducing Agile Development into Bioinformatics: An Experience
  Report}.
\newblock In {\em Agile Development Conference 2005}, pages 132--139, 2005.

\bibitem{kane2006}
David Kane, Moses Hohman, Ethan Cerami, Michael McCormick, Karl Kuhlmman, and
  Jeff Byrd.
\newblock {Agile Methods in Biomedical Software Development: a Multi-Site
  Experience Report}.
\newblock {\em BMC Bioinformatics}, 7(1):273, 2006.

\bibitem{kelly2009}
Diane Kelly, Daniel Hook, and Rebecca Sanders.
\newblock {Five Recommended Practices for Computational Scientists Who Write
  Software}.
\newblock {\em Computing in Science \& Engineering}, 11(5):48--53, 2009.

\bibitem{kelly2008}
Diane Kelly and Rebecca Sanders.
\newblock {Assessing the Quality of Scientific Software}.
\newblock In {\em First International Workshop on Software Engineering for
  Computational Science and Engineering}, May 2008.

\bibitem{kelt2008}
Douglas~A. Kelt, James~A. Wilson, Eddy~S. Konno, Jessica~D. Braswell, and
  Douglas Deutschman.
\newblock {Differential Responses of Two Species of Kangaroo Rat
  (\emph{Dipodomys}) to Heavy Rains: A Humbling Reappraisal}.
\newblock {\em Journal of Mammalogy}, 89(1):252--254, 2008.

\bibitem{killcoyne2009}
Sarah Killcoyne and John Boyle.
\newblock {Managing Chaos: Lessons Learned Developing Software in the Life
  Sciences}.
\newblock {\em Computing in Science \& Engineering}, 11(6):20--29, 2009.

\bibitem{kniberg2007}
Henrik Kniberg.
\newblock {\em {Scrum and XP from the Trenches}}.
\newblock Lulu.com, 2007.

\bibitem{lees2007}
David~C. Lees and Robert~K. Colwell.
\newblock {A strong Madagascan rainforest MDE and no equatorward increase in
  species richness: re-analysis of 'The missing Madagascan mid-domain effect',
  by Kerr J.T., Perring M. \& Currie D.J. (Ecology Letters 9:149–159, 2006)}.
\newblock {\em Ecology Letters}, 10(9):E4--E8, 2007.

\bibitem{letovsky1986}
S.~Letovsky.
\newblock {Cognitive processes in program comprehension}.
\newblock In {\em Empirical Studies of Programmers}, pages 58--79, 1986.

\bibitem{ma2007}
Che Ma and Geoffrey Chang.
\newblock {Retraction for Ma and Chang, Structure of the multidrug resistance
  efflux transporter EmrE from \emph{Escherichia coli}}.
\newblock {\em Proceedings of the National Academy of Sciences}, 104(9):3668,
  2007.

\bibitem{martin2002}
Robert~C. Martin.
\newblock {\em {Agile Software Development, Principles, Patterns, and
  Practices}}.
\newblock Prentice Hall, 2002.

\bibitem{matthews2008}
David Matthews, Greg Wilson, and Steve Easterbrook.
\newblock {Configuration Management for Large-Scale Scientific Computing at the
  {UK Met Office}}.
\newblock {\em Computing in Science \& Engineering}, pages 56--64,
  November-December 2008.

\bibitem{mcconnell2004}
Steve McConnell.
\newblock {\em {Code Complete: A Practical Handbook of Software Construction}}.
\newblock Microsoft Press, 2 edition, 2004.

\bibitem{merali2010}
Zeeya Merali.
\newblock {Error: Why Scientific Programming Does Not Compute}.
\newblock {\em Nature}, 467:775--777, 2010.

\bibitem{meszaros2007}
Gerard Meszaros.
\newblock {\em {xUnit Test Patterns: Refactoring Test Code}}.
\newblock Addison-Wesley, 2007.

\bibitem{noble2009}
William~Stafford Noble.
\newblock {A Quick Guide to Organizing Computational Biology Projects}.
\newblock {\em PLoS Computational Biology}, 5(7), 2009.

\bibitem{oberkampf2010}
William~L. Oberkampf and Christopher~J. Roy.
\newblock {\em {Verification and Validation in Scientific Computing}}.
\newblock Cambridge University Press, 2010.

\bibitem{openprovenance}
{Open Provenance}.
\newblock http://openprovenance.org.
\newblock Viewed June 2012.

\bibitem{oram2010}
Andy Oram and Greg Wilson, editors.
\newblock {\em Making Software: What Really Works, and Why We Believe It}.
\newblock O'Reilly, 2010.

\bibitem{osherove2009}
Roy Osherove.
\newblock {\em {The Art of Unit Testing: With Examples in .NET}}.
\newblock Manning, 2009.

\bibitem{perez2007}
Fernando P\'erez and Brian~E. Granger.
\newblock {IP}ython: a {S}ystem for {I}nteractive {S}cientific {C}omputing.
\newblock {\em {C}omput. {S}ci. {E}ng.}, 9(3):21--29, May 2007.

\bibitem{pitt-francis2008}
Joe Pitt-Francis, Miguel~O. Bernabeu, Jonathan Cooper, Alan Garny, Lee
  Momtahan, James Osborne, Pras Pathmanathan, Blanca Rodriguez, Jonathan~P.
  Whiteley, and David~J. Gavaghan.
\newblock {Chaste: Using Agile Programming Techniques to Develop Computational
  Biology Software}.
\newblock {\em Philosophical Transactions of the Royal Society A: Mathematical,
  Physical and Engineering Sciences}, 366(1878):3111--3136, September 2008.

\bibitem{pouillon2010}
Yann Pouillon, Jean-Michel Beuken, Thierry Deutsch, Marc Torrent, and Xavier
  Gonze.
\newblock {Organizing Software Growth and Distributed Development: The Case of
  {Abinit}}.
\newblock {\em Computing in Science \& Engineering}, 13(1):62--69, 2011.

\bibitem{prabhu2011}
Prakash Prabhu, Thomas~B. Jablin, Arun Raman, Yun Zhang, Jialu Huang, Hanjun
  Kim, Nick~P. Johnson, Feng Liu, Soumyadeep Ghosh, Stephen Beard, Taewook Oh,
  Matthew Zoufaly, David Walker, and David~I. August.
\newblock {A Survey of the Practice of Computational Science}.
\newblock In {\em 24th ACM/IEEE Conference on High Performance Computing,
  Networking, Storage and Analysis}, pages 19:1--19:12, 2011.

\bibitem{prechelt2010}
Lutz Prechelt.
\newblock {Two Comparisons of Programming Languages}.
\newblock In Andy Oram and Greg Wilson, editors, {\em Making Software: What
  Really Works, and Why We Believe It}, pages 239--258. O'Reilly, 2010.

\bibitem{ray2009}
Deborah~S. Ray and Eric~J. Ray.
\newblock {\em {Unix and Linux: Visual QuickStart Guide}}.
\newblock Peachpit Press, 4 edition, 2009.

\bibitem{robinson2005}
Evan Robinson.
\newblock {Why Crunch Mode Doesn't Work: Six Lessons}.
\newblock http://www.igda.org/why-crunch-modes-doesnt-work-six-lessons, 2005.
\newblock Viewed Sept. 2013.

\bibitem{sanders2008}
R.~Sanders and D.~Kelly.
\newblock {Dealing with Risk in Scientific Software Development}.
\newblock {\em IEEE Software}, 25(4):21--28, July-August 2008.

\bibitem{segal2008a}
J.~Segal.
\newblock {Models of Scientific Software Development}.
\newblock In {\em First International Workshop on Software Engineering for
  Computational Science and Engineering}, 2008.

\bibitem{segal2008b}
J.~Segal and C.~Morris.
\newblock {Developing Scientific Software}.
\newblock {\em IEEE Software}, 25(4):18--20, 2008.

\bibitem{segal2005}
Judith Segal.
\newblock {When Software Engineers Met Research Scientists: A Case Study}.
\newblock {\em Empirical Software Engineering}, 10(4):517--536, 2005.

\bibitem{smith2011}
Peter Smith.
\newblock {\em {Software Build Systems: Principles and Experience}}.
\newblock Addison-Wesley, 2011.

\bibitem{spolsky2000}
Joel Spolsky.
\newblock {The Joel Test: 12 Steps to Better Code}.
\newblock http://www.joelonsoftware.com/articles/fog0000000043.html, 2000.
\newblock Viewed June 2012.

\bibitem{vardi2010}
Moshe Vardi.
\newblock {Science Has Only Two Legs}.
\newblock {\em Communications of the ACM}, 53(9):5, September 2010.

\bibitem{williams2010}
Laurie Williams.
\newblock {Pair Programming}.
\newblock In Andy Oram and Greg Wilson, editors, {\em Making Software: What
  Really Works, and Why We Believe It}, pages 311--322. O'Reilly, 2010.

\bibitem{wilson2006b}
Greg Wilson.
\newblock {Software Carpentry: Getting Scientists to Write Better Code by
  Making Them More Productive}.
\newblock {\em Computing in Science \& Engineering}, pages 66--69,
  November-December 2006.

\bibitem{wilson2013}
Greg Wilson.
\newblock {Software Carpentry: Lessons Learned}.
\newblock http://arxiv.org/abs/1307.5448, 2013.

\bibitem{xie2013knitr}
Yihui Xie.
\newblock knitr: A general-purpose package for dynamic report generation in r.
\newblock {\em R package version}, 1, 2013.

\bibitem{xunit}
{List of unit testing frameworks}.
\newblock http://en.wikipedia.org/wiki/List\_of\_unit\_\-testing\_frameworks.
\newblock Viewed June 2012.

\bibitem{zeller2009}
Andreas Zeller.
\newblock {\em {Why Programs Fail: A Guide to Systematic Debugging}}.
\newblock Morgan Kaufmann, 2009.

\end{thebibliography}

\pagebreak

\section*{Reference}

{\footnotesize
\begin{enumerate}

\item Write programs for people, not computers.
  \begin{enumerate}
  \item A program should not require its readers to hold more than a handful of facts in memory at once.
  \item Make names consistent, distinctive, and meaningful.
  \item Make code style and formatting consistent.
  \end{enumerate}

\item Let the computer do the work.
  \begin{enumerate}
  \item Make the computer repeat tasks.
  \item Save recent commands in a file for re-use.
  \item Use a build tool to automate workflows.
  \end{enumerate}

\item Make incremental changes.
  \begin{enumerate}
  \item Work in small steps with frequent feedback and course correction.
  \item Use a version control system.
  \item Put everything that has been created manually in version control.
  \end{enumerate}

\item Don't repeat yourself (or others).
  \begin{enumerate}
  \item Every piece of data must have a single authoritative representation in the system.
  \item Modularize code rather than copying and pasting.
  \item Re-use code instead of rewriting it.
  \end{enumerate}

\item Plan for mistakes.
  \begin{enumerate}
  \item Add assertions to programs to check their operation.
  \item Use an off-the-shelf unit testing library.
  \item Turn bugs into test cases.
  \item Use a symbolic debugger.
  \end{enumerate}

\item Optimize software only after it works correctly.
  \begin{enumerate}
  \item Use a profiler to identify bottlenecks.
  \item Write code in the highest-level language possible.
  \end{enumerate}

\item Document design and purpose, not mechanics.
  \begin{enumerate}
  \item Document interfaces and reasons, not implementations.
  \item Refactor code in preference to explaining how it works.
  \item Embed the documentation for a piece of software in that software.
  \end{enumerate}

\item Collaborate.
  \begin{enumerate}
  \item Use pre-merge code reviews.
  \item Use pair programming when bringing someone new up to speed and when tackling particularly tricky problems.
  \item Use an issue tracking tool.
  \end{enumerate}

\end{enumerate}
}

\end{document}